# Criminal Fishing System Based on Wireless Local Area Network Access Points
## —Can Media Access Control address assist criminal investigation?

Hiroaki Togashi, *Member, IEEE*, Yasuaki Koga, and Hiroshi Furukawa, *Member, IEEE*

*Abstract*—Currently, many Wi-Fi access points are being installed in urban areas. This paper considers how this infrastructure can be used to assist criminal investigations and improve public safety. We propose a criminal investigation assistance system that uses multiple wireless local area network (LAN) access points and cameras. The proposed "Criminal Fishing System" enumerates candidate media access control (MAC) addresses of culprits' mobile devices from probe request signals gathered by access points during the period in which a culprit is near the scene of an incident. Preliminary experiments demonstrated that the proposed system could identify the MAC address of the culprit's device, which would allow authorities to capture the culprit's radiowave fingerprint. After enumerating the candidate MAC addresses, the culprit's usual appearance can be obtained by surveilling these MAC addresses, especially when it changes less frequently. Moreover, the MAC address itself can be admissible as evidence that the culprit was near the scene of an incident, given that the MAC address is static, that is, it has not changed after the incident, or the original MAC address can be retrieved from the randomized MAC address.

## I. Introduction

CURRENTLY, the importance of traffic offloading from cellular networks to Wi-Fi networks is increasing owing to the rapid growth of network traffic [1] caused by the proliferation of smartphones and tablet devices. To extend Wi-Fi coverage, many Wi-Fi access points (APs) are being installed in urban areas.

Criminal activities, such as graffiti, shoplifting, larceny, and kidnapping, occur in urban areas. These incidents cause public expenditure and social damage. For example, the total cost for of cleaning up graffiti is approximately 12 billion USD per year in the United States [2]. The costs due to shoplifting are 461.5 billion yen per year in Japan [3] and 44 billion USD per year in the United States [4].

Against this backdrop, several crime prevention systems have been developed. Security cameras that capture photos or videos are widely used as security equipment. However, installation costs tend to be high, particularly when several cameras are installed and connected to a network. Another typical problem with security cameras is that a culprit whose face is hidden cannot be identified from images. This issue cannot be resolved by placing many cameras in a surveillance area.

We consider that using many APs in urban areas as a component of a criminal investigation assistance system can reduce the incidence of criminal activity and improve public safety. Among the several types of signals transmitted from a mobile device, some signals contain device identifiers. For example, a probe request signal is a signal broadcast by a Wi-Fi device to find surrounding Wi-Fi access points (APs), and it contains the media access control (MAC) address unique to each network interface. This paper describes a criminal investigation assistance system that utilizes the probe request signals gathered by Wi-Fi APs.

The remainder of this paper is organized as follows. Section 2 describes the research background, including brief introductions of related studies. Section 3 describes the proposed "Criminal Fishing System," and Section 4 describes a preliminary experiment performed using the proposed system. Section 5 discusses several associated issues, including ethical and privacy issues. Section 6 presents our concluding remarks.

## II. Research Background

### A. Existing crime prevention/detection systems

Security cameras are widely used as anti-crime equipment that record incidents by storing images captured from the scene of an incident. They are typically used as part of a closed-circuit television (CCTV) system. This system consists of cameras and monitors, and all pieces of equipment are connected locally, that is, the captured images are not broadcast publicly. However, the cost of installing security cameras is high, especially when multiple cameras with wired connections are to be installed. Another typical issue with security cameras is that if culprits hide their face, they cannot be identified from the captured images. This issue cannot simply be solved by installing many security cameras in a surveillance area.

Park et al. proposed a method [5] that recognizes people's emotions by using wearable sensors, namely, heartbeat sensors and temperature sensors. By combining data from these sensors with CCTV images, the method detects criminal behaviors. Its detection results are stored in a database. iProtect [6] is a method based on activity recognition that utilizes sensors equipped on a target user's smartphone. The method involves anomaly detection; it detects differences between criminal behaviors, for instance, assaults, and routine activities. However, for detecting criminal behaviors, these methods require the attachment of sensing equipment on the target person or the use of special software on their smartphones. Therefore, these methods are not practical because culprits disable crime-detection software or equipment before committing a crime. We aim to achieve criminal fishing when the smartphone of a target person is simply turned on.

*B. Probe Request signal*

Probe request signals are broadcast when a Wi-Fi device searches APs or requests a connection to an AP which has the extended service set identifier (ESS-ID) stored as the preferred network list (PNL) in the device. There are two methods of establishing connections between Wi-Fi devices and APs; Active scan, in which a device broadcasts probe request signals to the surrounding APs and receives probe responses from the APs. Thereafter, procedures to establish a connection between them are executed. In passive scan, a beacon signal is broadcast periodically by APs to establish network connections. Most smartphones broadcast probe request signals at random intervals to execute active scan and swiftly establish network connections. The time interval with which a Wi-Fi device broadcasts probe request signals differs according to the device type, its connection state, operating system, and software running on the device. Generally, this interval lasts from a few seconds to a few minutes. A device tends to broadcast probe request signals more frequently when it is not connected to any AP and its display is on.

The following information can be obtained by analyzing probe request signals.
- Sender MAC address
- RSSI
- ESS-ID (included only when declared explicitly)

Currently, about 3.4 billion smartphones are in use worldwide [1], and each device broadcasts probe request signals when its Wi-Fi function is active. We call this signal as "radiowave fingerprint" because it contains the MAC address unique to each device. This "radiowave fingerprint" can be used to devise a novel security surveillance system by combining APs capable of capturing probe request signals and web cameras attached onto APs.

*C. MAC spoofing/randomization*

MAC addresses are factory-assigned and unique to each network interface. MAC spoofing is a software-based technique to change this MAC address. To spoof a MAC address, a certain level of technical skill is required. In the case of a sudden or minor criminal incident, the culprit is less likely to spoof the MAC address of their Wi-Fi device. Notably, to spoof a MAC address, most mobile devices also require "rooting" or "jail-breaking," which would make the device vulnerable to cyber attacks. In case that the spoofed MAC address is changed infrequently, the mobile device can still be tracked in a manner similar to a device with the factory-assigned (not-spoofed) MAC address. Therefore, MAC address spoofing itself is not a very efficient counter measure against mobile device tracking.

Recently, several operating systems (OSs) have been equipped with MAC address randomization. For example, iOS 10 uses a randomized MAC address for Wi-Fi scanning most of the time [7]. The latest version of Android OS and Windows OS are also equipped with MAC address randomization, but very few network modules and drivers are capable of randomizing MAC addresses; most Android/Windows devices cannot randomize MAC addresses. Recent research [8] has revealed that current MAC randomization is not effective to avoid mobile device tracking. The best way for a person to avoid being tracked when using Wi-Fi signals is to turn off their mobile devices.

In this paper, we implement a prototype of the proposed system and conduct a preliminary experiment without considering issues pertaining to MAC spoofing and randomization. Discussions related to these technologies are given later in the paper.

III. CRIMINAL FISHING SYSTEM

The proposed Criminal Fishing System finds clues to identify culprits by their "radiowave fingerprint," in addition to their camera images. Note that the MAC address is unique to each mobile device (network interface), not to each person, even if the address is factory-assigned. Therefore, aim of this system is not to identify a culprit but to find a clue to identify culprits. As mentioned in Sec. 2.C, there exist several techniques to hide the factory-assigned MAC address. However, if the candidate MAC addresses of the culprit's device can be enumerated, we consider that this information can help with criminal investigation as follows:

1) When the MAC address of a certain person's mobile device is included in the list of candidate MAC addresses, this points to the likelihood of that person being the culprit, an accomplice, eyewitness, or a victim of the incident.

2) When candidate MAC addresses are enumerated, the device owner's usual appearance can be obtained by monitoring the candidate addresses in several surveillance areas. In the case that the person's appearance is similar to the culprit's appearance, the person is likely to be the culprit of the incident. In the case that the appearance has little in common with the culprit's appearance seen around the scene of the incident, the MAC address must be regarded as that of a bystander's device.

Intended users and incident types of the proposed system are as follows.

Intended users:
• Facility administrators (when the surveillance area consists of private buildings, e.g., shopping malls, company office)
• Local police or security companies (when the area is public, e.g., station square, park)

Intended incident types:
• Personal or property crimes (e.g., shoplifting, graffiti, larceny)

*A. System outline*

This system employs PCWL-0200 [9] as APs to gather "radiowave fingerprints" and a USB camera mounted on each AP to shoot images around it. These radiowave fingerprints and images are transmitted to an administrative server along with an attached timestamp. When an incident occurs, the user of the Criminal Fishing System, for example, a facility administrator, checks the images to determine the duration of an incident. The system filters radiowave fingerprints observed during the period and enumerates candidate radiowave fingerprints of the culprit's device from a large number of obtained fingerprints. When the enumerated radiowave fingerprints are detected in another area, the appearance and the face of the device owner, that is, the candidate culprit can be recognized by camera images. This system can help accelerate crime investigations.

Even if the detailed appearance of a culprit cannot be obtained around the scene of an incident, it can be obtained later in another place by surveilling the enumerated MAC addresses, namely, MAC address contained in the radiowave fingerprints, in several areas. When the MAC address is captured in an area, the appearance of the device owner can be obtained from camera image shot around there. The probability of identifying culprits can be improved by locating APs over wide areas with high density.

*B. System components*

Figure 1 shows the components of the system. The main components are administrative server, PCWL-0200 (Wi-Fi AP), and USB camera. Each AP gathers probe request signals, and periodically transmits "radiowave fingerprints" to the administrative server. Each AP executes "motion," an image-capturing program that captures images when the program recognizes moving objects. These images are transmitted to the administrative server via a wireless backhaul network. The transmitted radiowave fingerprints and images are stored in the database on the server. To determine the culprit's device from a large volume of stored data, the administrative server is equipped with a web application. A facility administrator uses the web application to determine the duration for which the culprit stays around the scene of the incident by checking the stored images. The administrative server enumerates candidate MAC addresses of the culprit's mobile device based on this duration.

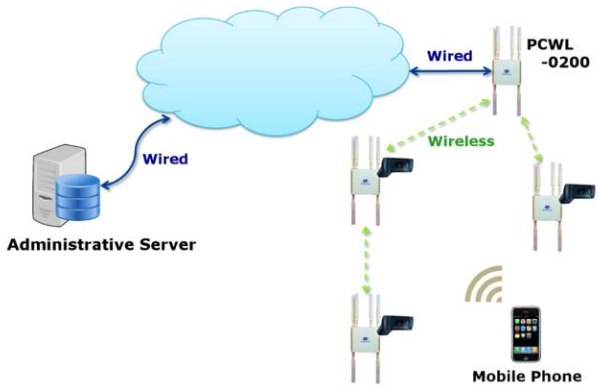

Fig. 1. Criminal fishing system. Each access point gathers probe request signals broadcast by surrounding devices. These signals and images captured around each AP are transmitted to the administrative server via a wireless backhaul network.

*C. Culprit determination algorithm*
*1) Outline of algorithm*

Because each AP gathers probe request signals from multiple devices, including the culprit's device and bystanders' devices, the MAC addresses of the bystanders' devices should be eliminated. These devices are roughly classified into the following 3 categories; "stable devices" are devices that stay at the same place, "long-distance devices" are devices that are located at some distance from the camera, and "short-distance devices" are devices that are located within a short distance from the camera.

The MAC addresses of the stable devices are always observed by a certain AP regardless of time. By contrast, the device of the culprit eventually moves away from the scene of the incident, and its MAC address is not observed by the AP after a while. Therefore, stable devices can be eliminated by determining the duration for which the culprit stays around the scene of the incident from camera images and eliminating the MAC addresses observed after the culprit has fled.

The long-distance devices are located far from the AP, and the RSSI of the signals emitted from these devices will be lower than that of the culprit's device. Therefore, these devices can be eliminated by setting a lower limit threshold on the RSSI of the probe request signals.

Short-distance devices can be categorized into "partially short-distance devices," the device of the person that behaves different to the culprit in the area of other APs, and "fully short-distance devices," the device of the person that behaves similarly to the culprit in the area of other APs. The partially short-distance devices can be eliminated by extracting MAC addresses corresponding to the time when the culprit stays around each AP distributed around the area and by eliminating the MAC addresses that are not commonly observed by all target APs. In case of the fully short-distance devices, it is considered not necessary to eliminate them because these radiowave fingerprints indicate the devices of accomplices or those of the victims of the incident, which are useful for determining the culprit. In case that the addresses of entirely unrelated persons exist, they can be eliminated as follows. When the MAC addresses were observed at a place, their appearance can be seen by checking the images shot at the place. If they can be recognized as unrelated persons by checking

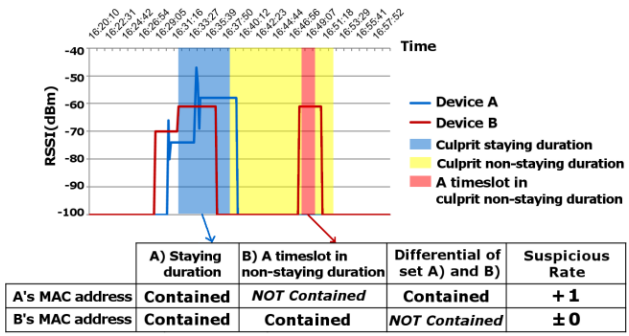

Fig. 2. Calculation of suspicious rate. The suspicious rate is the index showing the probability that a given MAC address is the culprit's address. It is calculated based on the appearance of each MAC address around the time of a given incident.

images taken around other APs, their address can be eliminated from the candidate list.

In terms of the filtering method, we need to discuss two performance indices: culprit detection rate and misdetection rate of other persons. The occurrence of a misdetection means that an innocent bystander might be treated as a culprit. This would increase the psychological resistance to implementation of the proposed system. It is essential to get a nod from community residents before introducing the system for practical use. Therefore, a low misidentification rate should be achieved, in addition to achieving high identification rate of the culprit. Based on this concept, the linear weighting method is developed.

*2) Linear weighted filtering*

"Linear weighted filtering" is a method to obtain candidate MAC addresses of a culprit's mobile device from a multitude of probe request signals. Linear weighted filtering uses the "suspicious rate," an index that shows the probability that a given MAC address is the one of the culprit's device. The suspicious rate is calculated for each MAC address based on the time at which the address was observed.

Linear weighted filtering calculates the difference among the MAC addresses observed during the "culprit staying duration" and that observed during the "culprit non-staying duration" (Fig. 2). These durations are defined by determining the times at which a culprit enters and exits the vicinity of a given incident scene. The stored photos are used to determine the durations. Because the addresses included in the differential set are likely to be the culprit's, the filtering method increments their suspicious rate. This method divides the culprit non-staying duration into intervals 30 s length and performs this calculation for each time slot.

The probability that the culprit's fingerprint will be collected by an AP around an incident scene is expected to decrease relative to the time elapsed from which the culprit was recognized as being in the given location. Therefore, the accuracy of culprit identification can be improved by increasing the value to be incremented by the "suspicious rate" relative to the time elapsed since the culprit staying duration, as follows. In this formula, n indicates the number of time slots, numbered from each end of the culprit' staying duration.

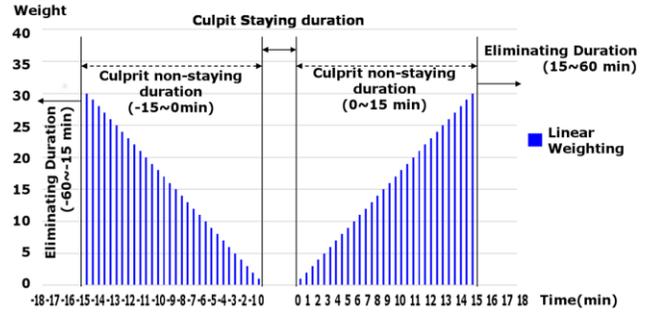

Fig. 3. Change in linear weighting over time. The value is increased according to the time elapsed since the culprit staying duration.

$$linear\_weighting(n) = \frac{2}{31}(n + 1)$$

We refer to this weighting method as "linear weighting" because the value is linearly increased according to the time elapsed since the culprit's staying duration, as shown in Fig. 3.

The addresses with the maximum RSSI values during the culprit staying duration and the maximum suspicious rates greater than the threshold are extracted as candidate MAC addresses. The filtering results are presented in a table that consists of a MAC address, its suspicious rate for each AP, and sum of the rates. When several addresses appear in the candidate list, an administrator can determine the MAC address

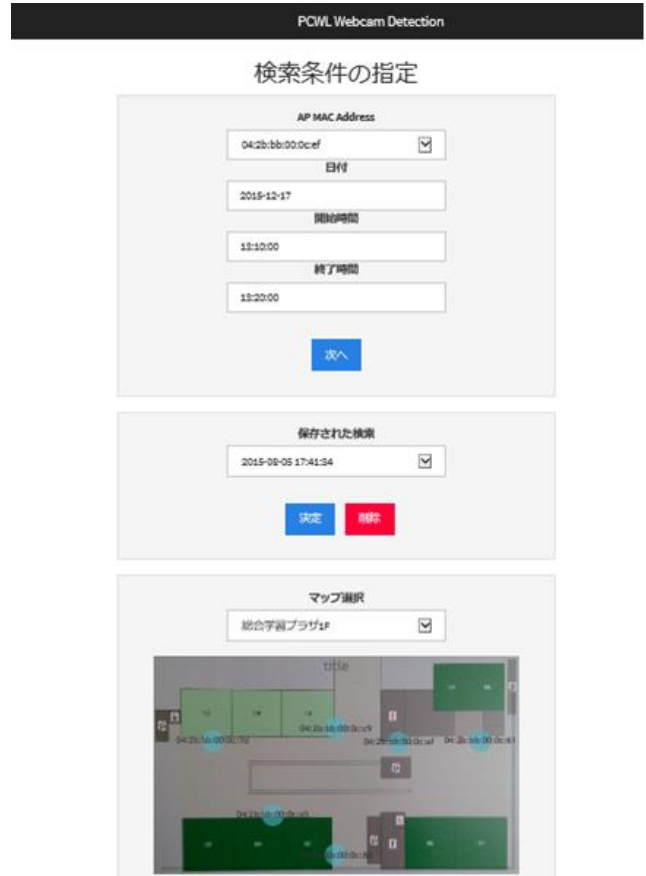

Fig. 4. Query input screen of Criminal Fishing web application. Facility administrator inputs surveillance area, AP identifier, date, and approximate time upon acknowledging the occurrence of a criminal incident.

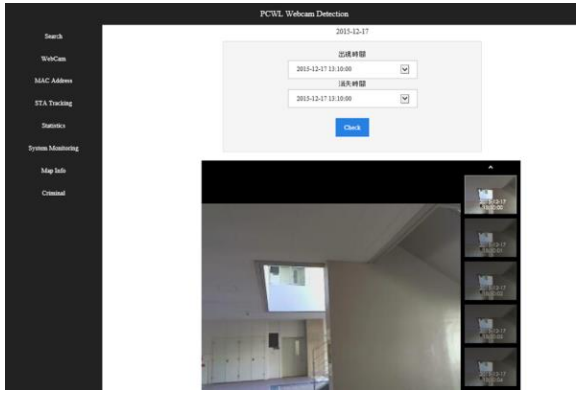

Fig. 5. List view of candidate photos. Detailed culprit staying duration at a certain AP is determined by choosing an appropriate photo.

that most likely belongs to the culprit by referring to the suspicious rates.

*D. Criminal Fishing web application*

This system collects many photos shot around several APs. To find appropriate photos for determining the culprit staying duration, we developed the Criminal Fishing web application. This application shows candidate photos by inputting surveillance area, AP identifier, date, and approximate time (Fig. 4). Detailed culprit staying duration is determined by choosing an appropriate photo from the candidate photos shown in the list view (Fig. 5). The Criminal Fishing System extracts candidate MAC addresses from the culprit staying duration. By executing the same procedures on each AP that the culprit has passed in a certain surveillance area, candidate MAC address are output as the filtering result.

## IV. PRELIMINARY EXPERIMENT

A preliminary experiment was conducted to examine whether the proposed system can enumerate the MAC address of the culprit's device from a list of candidate addresses. This experiment was performed in an experimental environment, in which a culprit held a mobile device that broadcast probe request signals frequently with the factory-assigned MAC address.

*A. Experimental settings*

In this experiment, a surveillance area was set up with six APs, as shown in Fig. 6. Specifically, one AP called the "core" was used to aggregate data communication to the database server, and the other five APs called "slaves" captured probe request signals. Because the image-capturing program is computational resource-heavy, it may interrupt APs' communication functions. To prevent this, five single-board computers (four PandaBoards [10] and one Raspberry Pi [11]) were used to capture photos with the USB cameras. Each pair of single-board computer and slave AP was placed at a given location. A simulated culprit held a mobile device with functional Wi-Fi and walked along a predefined route that passed all slave APs, while trying to blend into the crowd. The experiment was performed 10 times using 10 different devices. We consider that this number of trials is adequate for a preliminary experiment. The aim is to evaluate the basic functionality of the proposed system, and further evaluation should be done in a more practical environment. The candidate MAC addresses of culprit's device were enumerated using the Criminal Fishing System web application.

TABLE I
SPECIFICATION OF THE EXPERIMENT

| Experiment place | Open Learning Plaza, Kyushu University Ito campus |
|---|---|
| equipment | PCWL-0200: 6 (1 for data communicationaggregation 5 for capturing Probe Request signals) Pandaboard: 4 Raspberry Pi: 1 USBcamera: 5 Data server (CentOS): 1 Wi-Fi devices: 10 |

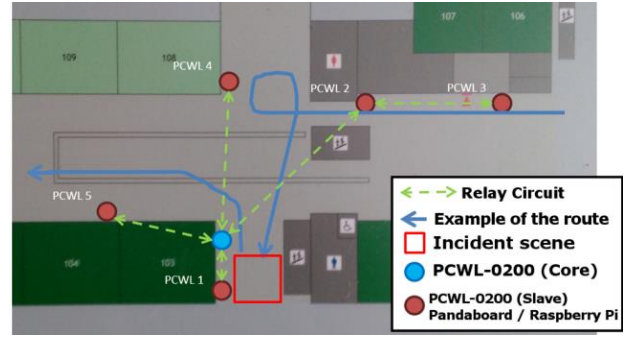

Fig. 6. Placement of APs in experiment. We use six Wi-Fi APs in the experiment; one AP is used for aggregating communication with the database server, and the other five capture probe request signals.

*B. Experimental results*

In this experiment, "radiowave fingerprints" of two (2 out of 10) devices were not observed on any AP, and not all single-board computers were able to capture images of the culprit. We believe that image capture failed because of the instability of the single-board computer, unstable signal strength, and so on. These problems can occur in more practical environments as well.

Table 2 lists the experimental results. The proposed method succeeded in enumerating the MAC address of the culprit's device 8 out of 8 times, that is, the culprit's mobile device broadcast probe request signals. In the fourth and the ninth trials, the proposed system was able to prevent false detection of the MAC address of a bystanders' device. However, because the culprit's device did not broadcast probe request signals, the system was not able to observe the MAC address of the culprit's device. "No. of APs used in identification" indicates that the number of the APs that were able to capture images. The maximum number was five in this experiment.

TABLE II
RESULTS OF EXPERIMENT

| Trial No. | No. of enumerated MAC/No. of all observed MACs | No. of APs used for identification | Identifiability of culprit |
|---|---|---|---|
| 1 | 1／60 | 4 | ○ |
| 2 | 1／54 | 5 | ○ |
| 3 | 1／62 | 3 | ○ |
| 4 | 0／31 | 4 | - |
| 5 | 1／41 | 5 | ○ |
| 6 | 1／42 | 4 | ○ |
| 7 | 1／40 | 3 | ○ |
| 8 | 1／35 | 2 | ○ |
| 9 | 0／35 | 3 | - |
| 10 | 1／54 | 2 | ○ |

The experimental results show that the proposed system can find the MAC address of the culprit's device by fusing probe request signals and camera images.

## V. DISCUSSION

### A. Ethical issues of proposed system

As mentioned in Sec. 3, the main purpose of the proposed system is to enumerate candidate MAC addresses of a culprit's mobile device. A MAC address by itself is not an identifier of a person. However, when a malicious user spies and obtains a target person's MAC address, the malicious user can track the target by using said address. To prevent the Criminal Fishing System from being accessed by unintended users, the criminal fishing server and the APs set up a surveillance area should be located on same and closed local area network.

In addition, we must deal with privacy issues. Although a MAC address is not a direct identifier of any person, it is correlated to a certain person strongly. In the case of factory-assigned MAC address, it is definitely unique to each device, and a mobile device travels along with its owner. If the mobile device can be tracked, the owner of the device can be tracked. Although MAC address is not regarded as personal information in most countries, we should notify that the Criminal Fishing System is in operation in a given surveillance area.

### B. Effects of MAC address spoofing and randomization

We should consider how the proposed system works if a culprit has a mobile device with a spoofed or randomized MAC address. In short, MAC spoofing and randomization have no critical effect on this system. However, these technologies sometimes trigger frequent changes of MAC address, which undeniably affects the proposed system. In the case that a MAC address is changed very frequently, for example, every few minutes, it cannot be enumerated as one of the candidate MAC addresses. However, according to the preliminary experiment, the proposed system can avoid enumerating the MAC addresses of bystanders' mobile devices. Therefore, it can prevent innocent bystanders from being entangled into a criminal investigation.

Depending on the frequency of MAC address change, the proposed system can assist criminal investigation as follows. If the MAC address is not changed in several days (may be several hours), the whereabouts of the device owner, that is, candidate culprit, can be obtained by monitoring the MAC address in several areas. In this manner, the system can assist criminal investigations, even if the MAC address is been changed before candidate culprit is questioned about the incident. In the case that the MAC address is rarely changed, regardless of whether it is factory-assigned or spoofed, this MAC address itself can be admissible as evidence that the person was present at the scene when the incident occurred.

Currently, MAC address randomization algorithms and their implementation differ across operating systems. A single standardized and secure randomization method has not been established. According to [8], several vulnerabilities remain in MAC address randomization, and some studies [12][13] have dealt with target device tracking regardless of whether the device randomizes its MAC address.

### C. Social effects of proposed system

This section discusses the social effects of the proposed system, independent of the MAC randomization or spoofing issues discussed in Sec. 5.B.

The experimental results demonstrated that the proposed system can find the MAC address of a culprit's mobile device accurately when the culprit has a device with active Wi-Fi. According to the literature [1], 3.4 billion smartphones are used actively worldwide, and mobile broadband subscriptions have reached 3.6 billion. In short, nearly half of the world's population uses devices that broadcast probe request signals. Based on the experimental results, the proposed system can identify the MAC addresses of approximately 50% of the people in the area covered by APs capable of capturing probe request signals. In urban areas, many Wi-Fi APs are being installed, and these APs will be replaced with more sophisticated APs in the near future. Therefore, the system is anticipated to be effective for approximately one-half of the criminal incidents that occur in urban areas. It will be a strong deterrent against criminal incidents in cities. Moreover, we believe that the proposed AP-based criminal prevention system is beneficial because culprits will be unaware of the locations of the APs and connection status of their smartphones when perpetrating criminal activities.

### D. Other issues for practical use

We consider that there remain two major issues before introducing the proposed system into practical use. The first is the failure to capture "radiowave fingerprint" from the culprit's device. In the experiment, the device held by the simulated culprit was set to emit probe request signals frequently to ensure that the APs captured the "radiowave fingerprint." Sometimes, however, some APs are able to capture the culprit's "radiowave fingerprint" and others are not in practical situations. Currently, the proposed method can identify MAC address of the culprit's device only when all APs that have captured the culprit's image succeed in capturing the signal.

The other issue is that the misdetection rate increases as the number of APs that can be used to identify the culprit's "radiowave fingerprint" decreases. In the experiments, the simulated culprit walked around the surveillance area passing many APs. However, a real culprit would not always move around the area while passing near the APs. The culprit identification rate of the proposed system is expected to decrease in such situations. By contrast, the number of APs used for culprit identification can be used as an index to measure the reliability of the identification result. We believe that they can help innocent bystanders from being entangled in criminal investigations, not only by referring to the detected results of the proposed system but also considering the number of APs that captured the culprit's "radiowave fingerprint."

## VI. Conclusion

We presented the Criminal Fishing System, which uses radiowave fingerprints gathered by densely placed APs. Specifically, this system enumerates candidate MAC addresses of a culprit's mobile device based on the suspicious rate, which is calculated on the basis of the appearance of each MAC address around the time of a given incident. A preliminary experiment demonstrated that the proposed system can find the MAC address of a culprit's device by fusing probe request signals and camera images. These results indicate that Wi-Fi based target tracking and camera-based monitoring have a certain level of utility in investigating crimes.

In the future, we plan to improve the proposed system by testing it in a more practical environment, modifying it for use in the context of specific types of incidents. The existing algorithm of the proposed system is not intended for any dedicated type of crimes. We consider that the accuracy of our system can be improved by considering criminal characteristics.

In addition, we would like to incorporate the anti-MAC spoofing/randomization techniques mentioned in Sec. 5 B into our proposed system and evaluate the effects of doing so on our Criminal Fishing System. We believe that accurate target tracking using Wi-Fi signals can improve the accuracy and the utility of the proposed system.

## References

[1] "Ericsson Mobility Report–Mobile World Congress edition, February 2016," [Online] Available: http://www.ericsson.com/res/docs/2016/mobility-report/ericsson-mobility-report-feb-2016-interim.pdf
[2] D. L. Weisel, "Problem-Oriented Guides for Police Series No. 9, Graffiti," 2002, [Online] Available: http://www.popcenter.org/problems/graffiti
[3] Japan Association of Electronic Article Surveillance Machines, "Electronic Article Surveillance System Handbook, 3rd. edition," 2012. (In Japanese)
[4] National Retail Federation, "National Retail Security Survey 2015," [Online] Available: https://nrf.com/resources/retail-library/national-retail-security-survey-2015
[5] J. Y. Byun, A. Nasridinov and Y. H. Park, "Internet of things for smart crime detection," Contemporary Engineering Sciences 7(13):749-754 January 2014
[6] Z. Zhenyu, H. Guo, and Y. E. Sun, "iProtect: Detecting Physical Assault Using Smartphone," 10th International Conference Wireless Algorithms, Systems, and Applications (WASA 2015), Qufu, China, August 10-12, 2015.
[7] Apple Inc., "iOS Security–iOS 10," [Online] Available: https://www.apple.com/business/docs/iOS_Security_Guide.pdf
[8] J. Martin et. al., "A Study of MAC Address Randomization in Mobile Devices and When it Fails," [Online] Available: https://arxiv.org/abs/1703.02874
[9] PicoCELA Inc., "PCWL-0200," [Online] Available: http://jp.picocela.com/12477.html (In Japanese)
[10] pandaboard.org, "PandaBoard," [Online] Available: http://pandaboard.org/
[11] Raspberry Pi Foundation, "Raspberry Pi," [Online] Available: https://www.raspberrypi.org/
[12] J. Pang et al., "802.11 user fingerprinting," Proceedings of the 13th annual ACM international conference on Mobile computing and networking (MobiCom'07), Montréal, Canada, Sep. 09 - 14, 2007 pp.99-110.
[13] M. Vanhoef et al., "Why MAC Address Randomization is not Enough: An Analysis of Wi-Fi Network Discovery Mechanisms," Proceedings of the 11th ACM on Asia Conference on Computer and Communications Security (ASIA CCS '16), Xi'an, China, May 30–June 3, 2016, pp.413-424.
[14] P. Robyns et al., "Non-cooperative 802.11 MAC layer fingerprinting and tracking of mobile devices," Security and Communication networks, 2017.

**Hiroaki Togashi** (M'16–) received a B.A. (Environmental Information) from Keio University in 2005, and Ph.D. from the Graduate University for Advanced Studies in 2012.

From 2013 to 2015, he worked as a technical staff and research scientist at the Center for Service Research, National Institute of Advanced Industrial Science and Technology (AIST). While there, he participated in studies on human sensing systems using smartphone sensors. The results have become part of a patent. Since 2015, he has been an Assistant Professor in the Faculty of Information Science and Electrical Engineering, Kyushu University. His current research interests include indoor positioning and "Internet of Things" (IoT) systems.

**Yasuaki Koga** received B.S. and M.S. degrees in Engineering from Kyushu University in 2016.

He is now with NISHIMU ELECTRONICS INDUSTRIES Co., LTD. He was with the Graduate School of Information Science and Electrical Engineering, Kyushu University, Fukuoka, Japan.

**Hiroshi Furukawa** (M' 94-) received a B.S. degree in Information Engineering from Kyushu Institute of Technology in 1992 and a Doctorate in Electrical and Electronic Engineering from Kyushu University in 1998.

From 1992 to 1996, he taught at the Department of Computer Science and Electronics, Kyushu Institute of Technology. From 1996 to 2003, he worked at Networking Research Laboratories, NEC Corporation, where he oversaw studies on radio resource management for 3G. While there, his patented Site Selection Diversity Transmit (SSDT), which became part of the IMT-2000 standard, including WCDMA and CDMA2000. In 2003, he left NEC to pursue his own research as an Associate Professor in the Department of Intelligent Systems at Kyushu University. Since 2010, he has been a Professor in the Department of Advanced Information Technology at the same university. Since he moved to Kyushu University, he has focused on studies on wireless backhaul technology for super-small cells, such as wireless LAN. He realized a wireless backhaul protocol suite for stable wireless multi-hop relay under dynamic radio conditions as a fruit of more than 10 years of study in this field. He also serves as a Representative Director and CTO of PicoCELA, Inc., a company tasked with commercialization of the aforementioned wireless backhaul technology. His patented wireless backhaul network system has been deployed in more than 100 sites in Japan, reducing 70% LAN cabling on average.

He is a member of IEEE and IEICE, and received the Young Engineer Award from IEICE in 1995. He has served as a TPC member in many journals and international conferences and as a committee member in national projects. He has been granted 50 patents worldwide and has published three books and more than 110 technical papers in journals, transactions, and international conferences.